\documentclass[
    ,final            % use final for the camera ready runs
  ]
  {aipproc}

\layoutstyle{8x11single}

\begin{document}

\title{Beyond the Standard Model of Cosmology}


\author{John Ellis}{
  address={CERN, 1211 Geneva 23, Switzerland}
}

\author{D.V. Nanopoulos}{
  address={George P. and Cynthia W. Mitchell Institute for Fundamental
Physics, Texas A\&M
University,\\ College Station, TX 77843, USA; \\
Astroparticle Physics Group, Houston
Advanced Research Center (HARC),
Mitchell Campus,
Woodlands, TX~77381, USA; \\
Academy of Athens,
Division of Natural Sciences, 28~Panepistimiou Avenue, Athens 10679,
Greece}
}

\begin{abstract}
Recent cosmological observations of unprecented accuracy, by WMAP in 
particular, have established a `Standard Model' of 
cosmology, just as LEP established the Standard Model of particle physics. 
Both Standard Models raise open questions whose answers are likely to be 
linked. The most fundamental problems in both particle physics and 
cosmology will be resolved only within a 
framework
for Quantum Gravity, for which the only game in 
town is string theory. We 
discuss novel ways to model cosmological inflation and late acceleration 
in a non-critical string approach, and discuss possible astrophysical 
tests.
\end{abstract}

\maketitle

%%%%%%%%%%%%%%%%%%%%%%%%%%%%%%%%%%%%%%%%%%%%
%% MAINMATTER
%%%%%%%%%%%%%%%%%%%%%%%%%%%%%%%%%%%%%%%%%%%%
\begin{center}
CERN-PH-TH/2004-219 ~~~~~~ ACT-06-04 ~~~~~~ MITP-04-21 ~~~~~~ 
astro-ph/0411153
\end{center}

\section{`D\'ej\`a vu' all over again?}

The current situation in cosmology has many striking parallels with the
situation that emerged in particle physics
a decade or so ago. Indications from many different types of astronomical
observations have converged on a
`Standard Model' of cosmology that is as successful as the Standard Model
of particle physics.

\begin{figure}[tbh]
  \resizebox{15pc}{!}{\includegraphics{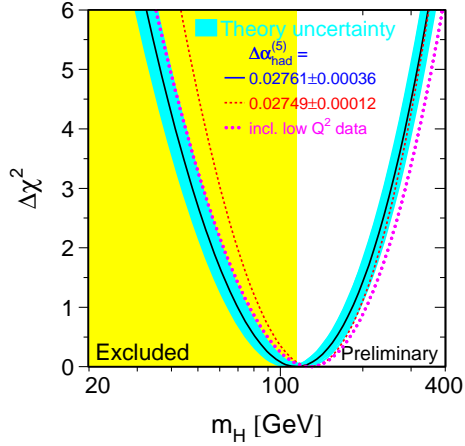}}
\caption{Data from LEP and other particle accelerators confirm predictions 
of the Standard Model with high accuracy, if there is a Higgs boson 
with mass is near the bottom of the $\Delta \chi^2$ curve indicated by 
the blue band~\protect\cite{LEPEWWG}. Direct searches at LEP have excluded 
a mass below 114~GeV, 
as shown by the yellow shading.} 
\label{fig:LEP} 
\end{figure}

In cosmology, the convergent indicators have included light-element
abundances, galactic rotation curves, COBE,
HST, large-scale structures, high-redshift supernovae, and many more
observations~\cite{Triangle}. In particle physics, the many
convergent experiments of the 1980's and before were validated and
confirmed much more precisely by measurements
at the LEP and other accelerators during the 1990's and 
subsequently~\cite{LEPEWWG}.
The missing link in the Standard Model of particle physics is the Higgs 
boson, and the current status of our 
information about it is shown in 
Fig.~\ref{fig:LEP}. Likewise, the `Standard Model' of
cosmology has been triumphantly
validated and confirmed this decade by the WMAP observations of the cosmic
microwave background (CMB)~\cite{WMAP}, but there are also some missing 
links, as 
discussed below.

\begin{figure}[tbh]
  \resizebox{15pc}{!}{\includegraphics{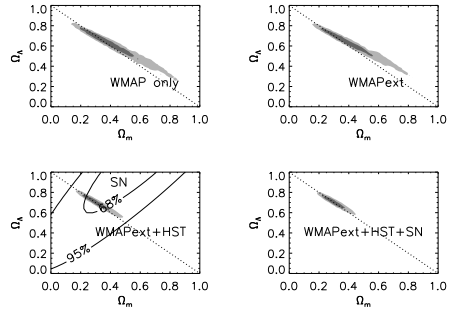}}
\caption{Data from WMAP, its extensions to include other CMB data, 
high-redshift supernovae and the HST all converge on a cosmology with 
$\Omega_m \sim 0.3, \Omega_\Lambda \sim 07$ and $\Omega_{tot} \sim 
1$~\protect\cite{WMAP}.} 
\label{fig:WMAP} 
\end{figure}

\begin{figure}[tbh]
  \resizebox{15pc}{!}{\includegraphics{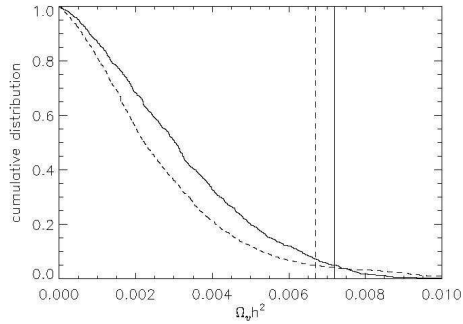}}
\caption{Data from WMAP, other CMB experiments and large-scale 
observations set an upper limit on $\Omega_\nu$ corresponding to an 
upper limit of 0.23~eV on each neutrino mass~\protect\cite{WMAP}.}
\label{fig:nu}
\end{figure}

WMAP is the LEP of cosmology. As seen in Fig.~\ref{fig:WMAP}, it has 
measured with greater precision than
was possible before the material and
energetic content of the Universe. In combination with other CMB 
measurements and large-scale
structure observations, WMAP has provided the
most stringent upper limit on the amount of hot dark matter and hence the
sum of the neutrino masses, as seen in Fig.~\ref{fig:nu}. WMAP
observations of the peaks in the CMB perturbation spectrum have confirmed
with higher precision the estimates of
the baryon density of the Universe made previously on the basis of
light-element abundances~\cite{BBN}. WMAP data confirm
that the total energy density of the Universe is very close to the critical
density, $\Omega_{tot} \sim 1$, and the combination of WMAP
with other CMB data favours an amount of cold dark matter $\Omega_m 
\sim 0.3$ that is
consistent with the previous indications from
large-scale structures and high-redshift supernovae~\cite{SN}. Combining 
all the
cosmological data, the amount of cold dark
matter is determined precisely and the need for substantial dark energy 
with $\Omega_\Lambda \sim 0.7$ becomes
overwhelming, as seen in Fig.~\ref{fig:WMAP}.

\section{That is not all!}

Despite its successes, this emergent `Standard Model' of cosmology is
deeply unsatisfactory. Just like the
Standard Model of particle physics, it fits (almost) all the data in its
respective field, but it is equally
incomplete, and the picture of the Universe that it provides is rather
bizarre. In particle physics, the
`standard' list of problems left unanswered by the Standard Model include
the origins of particle masses - which
are thought to be due to a Higgs boson - the proliferation of particle 
types
- which are thought to be related to
the matter-antimatter asymmetry observed in laboratory experiments - the
unification of particle interactions -
which seems to require a hierarchy of mass scales whose maintenance needs
some stabilizing mechanism, and the
formulation of a quantum theory of gravity. These problems find remarkable
parallels in issues beyond the
`Standard Model' of cosmology.

{\it What is the origin of inflation?} We are asked to believe that the
great size and age of the Universe are due
to an early phase of (near) exponential expansion, during which quantum
fluctuations in density laid the seeds for
the subsequent formation of structures in the Universe. The front-runner
for the inflationary mechanism is some
elementary scalar field (inflaton) analogous to the Higgs field of the 
Standard Model
of particle physics discussed above~\cite{Guth}, but other
possibilities may be suggested by string theory, as discussed below.

{\it What is the origin of matter?} This is generally thought to be linked
to the small asymmetry between matter
and antimatter seen in the laboratory, which is in turn thought to be
possible only because of the proliferation
of particle types. Possible microscopic mechanisms include decays of heavy
(s)neutrinos or CP-violating Higgs
physics.

{\it What is the origin of dark matter?} This cannot be composed of
conventional baryonic matter, nor of
neutrinos, and the favoured option is some sort of massive
weakly-interacting particle. A prime candidate is the
lightest supersymmetric particle (LSP)~\cite{LSP}: it turns out that the 
mass scale
required for a (formerly) thermal relic
dark matter particle coincides with the electroweak range required to
stabilize the mass hierarchy with
supersymmetry.

{\it What is the origin of the anti-hierarchy?} There is a remarkable
coincidence between the orders of magnitude
of the current densities of baryons, cold dark matter and dark energy. The
similarity of the baryon and cold dark
matter density might point towards some common origin. However, the
underlying physics appears to be very
different, and the proximity (within three orders of magnitude) of the
hadronic and electroweak scales is not
really sufficient to explain this first cosmic coincidence between the
visible and dark matter densities. (On the
other hand, other particle candidates for cold dark matter, such as 
axions or Wimpzillas, are less strongly
motivated and have less reason to acquire a relic density comparable to
conventional matter.) The second
coincidence, between the matter and dark energy densities, is even more
baffling. This is because the scaling with
time of the matter density is completely different from that expected for a
cosmological constant, so the present
similarity can only be temporary.

{\it What is the origin of dark energy?} Na\"ive models ascribe this to the
remnant density of some elementary
scalar field, possibly also analogous to the Higgs field. However, its
origin should properly be treated in the
context of a quantum theory of gravity, which is the only suitable
framework for discussing vacuum energy. To make
life more difficult, the present value of the dark energy density is not in
the ballpark predicted by previous
theoretical models, indeed, it is many orders of magnitude smaller than
identifiable contributions to the vacuum
energy from QCD, the electroweak Higgs field or supersymmetry breaking. 
This has motivated models in which the vacuum energy is relaxing towards 
zero, but cosmological data set increasingly stringent limits on its 
equation of state.

{\it What is the origin of the Big Bang?} This is perhaps the ultimate
challenge for any reputable quantum theory of
gravity. The known laws of physics break down when particle energies
approach and exceed $10^{19}$ GeV, which is
expected na\"ively to have been the case when the Universe was less than
about $10^{-43} s$ old. Even our current
understanding of string theory seems inadequate under the hot and dense
conditions so early in the Universe, and
so far one can only speculate about physics at the origin of the Big Bang
or before.

\section{The LHC of the Sky?}

Many of the outstanding problems in both physics and cosmology, beyond
their respective Standard Models, will be
addressed by the LHC on the one hand and the Planck Surveyor on the other.
The `expected' discoveries in the two
fields will impact directly each other's problems.

\begin{figure}[tbh]
  \resizebox{15pc}{!}{\includegraphics{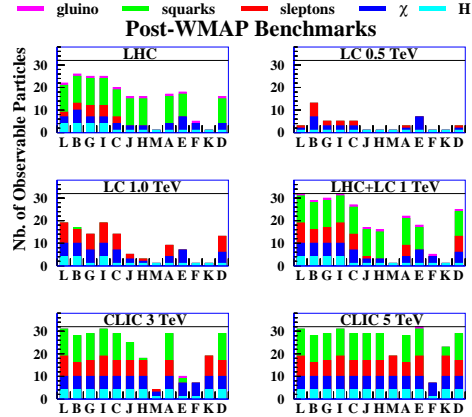}}
\caption{In many specific benchmark supersymmetric models that yield 
a relic dark matter density consistent with WMAP, labelled by 
capital letters, the LHC would observe many types of sparticles, and 
linear colliders would complement the information obtained from the 
LHC~\protect\cite{newbench}.} 
\label{fig:Bench} \end{figure}

\begin{figure}[tbh]
  \resizebox{15pc}{!}{\includegraphics{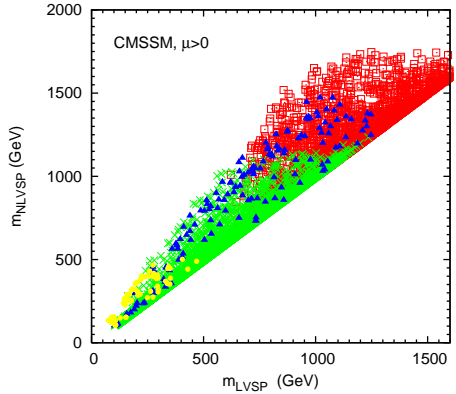}}
\caption{In a general sampling of supersymmetric models (red squares) the 
LHC (green crosses) would cover most of the region favoured by WMAP (blue 
triangles) and more than would be accessible to direct searches for dark 
matter scattering (yellow points)~\protect\cite{Tibet}.} 
\label{fig:CMSSM} 
\end{figure}

\begin{figure}[tbh]
  \resizebox{15pc}{!}{\includegraphics{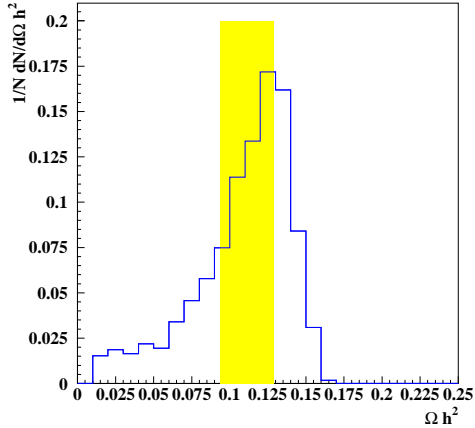}}
\caption{In benchmark model B of Fig.~\protect\ref{fig:Bench}, LHC 
measurements of the supersymmetric mass perimeters would enable the dark 
matter density to be calculated (histogram) with an accuracy comparable 
to the WMAP range (yellow band)~\protect\cite{newbench}.} 
\label{fig:Omega} 
\end{figure}

Simulations indicate that the LHC experiments ATLAS and CMS will be able to
detect or exclude a Standard
Model-like Higgs boson, whatever its mass~\cite{LHC}. The discovery of a 
Higgs boson
at the LHC would legitimize elementary
scalar fields and so give heart to their advocates as mechanisms for
inflation and dark energy. Likewise, as seen in 
Figs.~\ref{fig:Bench}, \ref{fig:CMSSM}, 
these
LHC experiments will be able to cover most of the space allowed for
supersymmetric models by cosmology. The
discovery of supersymmetry at the LHC would boost spectacularly the
candidacy of the LSP for cold dark matter,
whose density might (in favourable cases) be calculable using results from
LHC experiments, as seen in Fig.~\ref{fig:Omega}. Measurements of
CP-violating matter-antimatter asymmetries at the LHC (following those
already being measured at the B factories)
might cast light on the cosmological origin of matter. Any evidence of
extra dimensions would provide new cohesion
to the inchoate morass of string `phenomenology', with the potential to put
to experimental test models bearing
upon dark energy, inflation and the origin of the Big Bang itself.

What is the `LHC of the Universe', i.e. the
astronomical/astrophysical/cosmological instrument that will take us
beyond the `Standard Model' of cosmology? The most likely candidate is the
Planck Surveyor~\cite{PS} that is scheduled to be
launched in the same year as the startup of the LHC, namely 2007. It may be
able to take us into the area of
cosmological string phenomenology, for example by observing deviations from
the Gaussian, near-scale-invariant
adiabatic perturbations that are expected in conventional scalar-field
inflation. Or its measurements of
polarization in the CMB might provide a tell-tale signature of
gravitational waves. Certainly, we can expect it to
refine the present determinations of the conventional and dark matter
densities, providing a challenge for
microphysical measurements at the LHC and elsewhere.

\section{An End to `Field Theory as Usual'?}

It seems unlikely that Quantum Field Theory (QFT) will be challenged by the
particle physics beyond the Standard
Model that we expect to be discovered at the LHC, though this might be
possible in some scenarios for extra
dimensions that predict TeV-scale black holes, for example. Moreover, QFT
will probably also not be put into
question by many of the pressing cosmological problems, such as the origins
of conventional and dark matter.
However, some cosmological problems may find solutions only beyond the
conventional framework of QFT.

One glaring example is inflation. The CMB data from COBE to WMAP indicate
that it probably occurred at an energy
scale far beyond the electroweak scale, and potentially close enough to the
Planck scale for trans-Planckian
and/or stringy effects to be detectable. During the inflationary epoch, the
Universe is described by an
(approximately) de Sitter space, which has a horizon that provides it with
an effective `temperature' related to
the Hubble expansion rate. The Universe is described as a mixed quantum
state in which information is continually
being `lost' across the horizon, and must be summed over. The procedure for
doing this is well understood in
conventional QFT. However, as discussed above, inflation may be a
quantum-gravitational or stringy phenomenon, and
horizons pose conundra to string theorists.

String theory is formulated as a theory of the $S$ matrix. However, this
cannot be defined when mixed states enter
the game. Instead, scattering is formulated in terms of a \$ 
matrix~\cite{H,EHNS}. This
allows transitions between pure and mixed
states, or between different mixed states, that cannot be treated within
conventional $S$-matrix theory.

The enormity of this problem recently became apparent to string theorists
in the context of dark energy~\cite{Witten}. The fact
that the expansion of the Universe is currently accelerating
suggests that there is an analogous
horizon today and, unless the dark energy disappears in a rather abrupt
way, the asymptotic future state of the
Universe will continue to possess a horizon {\it ad infinitum}. This would
condemn string theorists to learn to
accommodate mixed states, generalize the $S$ matrix and embrace the \$ matrix.

Other considerations also motivate generalizing the $S$ matrix, notably the
horizons surrounding black holes, which
would seem to require a mixed-state description. String theorists have
noted triumphantly that the information
`lost' inside the horizon of a stringy black hole may in principle be
related to the properties of its internal
quantum microstates~\cite{EMNV}, raising the possibility of recovering the 
information
in the final stage of black-hole
evaporation. Also in principle, one could retain the $S$-matrix description
by keeping track of the quantum states
of such a stringy black hole. In practice, however, these are difficult to
observe, and one might necessarily be
reduced to summing over these string states and embracing the \$ matrix 
again~\cite{EMNQMV}.

\section{Do Not be so Critical}

\begin{figure}[tbh]
  \resizebox{15pc}{!}{\includegraphics{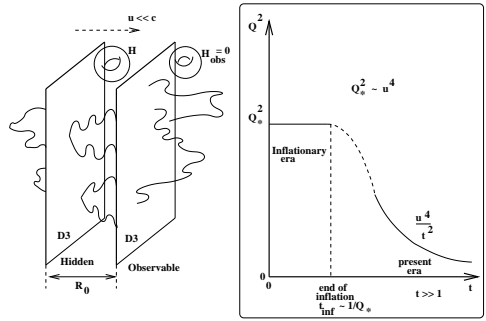}}
\caption{A specific non-critical string scenario for cosmological 
inflation, in which quantum effects in the presence of a pair of moving 
D-branes generates vacuum energy that may lead to near-exponential 
expansion~\protect\cite{EMNS2}.} 
\label{fig:EMNS}
\end{figure}

A plausible framework for addressing these issues is non-critical string
theory, in which one allows a deviation
from the critical value of the central charge normally imposed in string
theory. In the present context, the use
of non-critical strings for cosmology was first proposed as a natural way
to describe time-dependent space-time
backgrounds~\cite{ABEN}. Subsequently, this framework was also advocated 
for black-hole
physics, the non-criticality
reflecting the summation over unmeasured quantum states~\cite{EMNQMV}. 
This approach
provides a natural scenario for
cosmological inflation~\cite{EMNI}, in which the central-charge deficit is 
related
directly to the Hubble `constant' during
inflation, and reflects a summation over quantum states passing through the
de Sitter horizon. A specific
realization of this inflationary scenario has recently been 
proposed~\cite{EMNS2}, in
which moving D-branes generate the
departure from criticality and hence inflation, as illustrated in 
Fig.~\ref{fig:EMNS}. An analogous departure from criticality might also 
underlie the present apparent acceleration of the Universe~\cite{EMNA}.

Scattering in non-critical string theory does not admit an $S$-matrix
description, but instead requires use of the
\$ matrix. This may not suit critical string theorists, but may be forced
upon us by cosmology, as discussed above.
There has been a heated debate whether the \$ matrix is essential also at
the microphysical level, for example
because information may be continually lost across the mini-horizons of
Planck-scale black-hole fluctuations in
space-time foam. This fantastic possibility could be probed, for example,
in laboratory experiments on $K$ and $B$
mesons~\cite{EHNS,ELMN}. Other possible observable effects of space-time 
foam have been
proposed, such as an energy-dependent
modification of the velocity of photon propagation, which could perhaps be
tested by comparing the arrival times
of photons from pulsed astrophysical sources~\cite{Nature}. As shown in 
Fig.~{\ref{fig:GRB}, present data already tell us that any such effect 
can only be suppressed by some large energy scale close to the Planck 
mass or beyond~\cite{EMNS1}.

\begin{figure}[tbh]
  \resizebox{15pc}{!}{\includegraphics{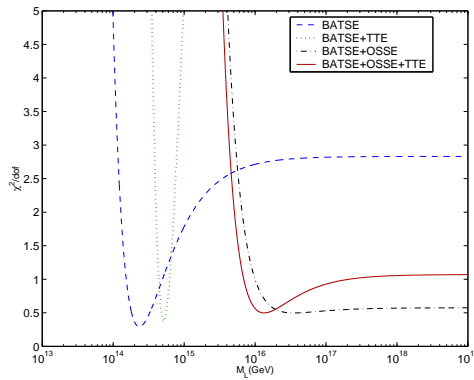}}
\caption{Upper limits on time-lags between photons of different energies 
coming from distant gamma-ray bursters set a lower limit $\sim 
10^{16}$~GeV on the scale of energy-dependent deviations from a universal 
velocity of light~\protect\cite{EMNS1}.} 
\label{fig:GRB} 
\end{figure} 
  
Quantum-gravity phenomenology is a very speculative subject, and the above
ideas may turn out to be incorrect
and/or untestable. However, if one is to find a convincing signature of
quantum gravity, it is surely better not
to just wait around for a burst of Hawking radiation from some
astrophysical black hole, but rather to search for
some characteristic phenomenon that goes beyond conventional QFT.

\section{Watch this Space}

This brief survey has highlighted some of the key aspects of both particle
physics and cosmology, emphasizing the important open questions in each
field, the prospects for resolving these problems, and the symbiosis
between the two fields. Cosmology and particle physics are inextricably
entangled, and solving their deepest problems may require a novel approach
to quantum gravity.


\begin{thebibliography}{9}

\bibitem{Triangle}
N.~A.~Bahcall, J.~P.~Ostriker, S.~Perlmutter and P.~J.~Steinhardt,
%``The Cosmic Triangle: Revealing the State of the Universe,''   
Science {\bf 284} (1999) 1481
[arXiv:astro-ph/9906463].
%%CITATION = ASTRO-PH 9906463;%%

\bibitem{LEPEWWG}
D.~Abbaneo {\it et al.} [LEP Electroweak Working Group],
{\tt http://lepewwg.web.cern.ch/LEPEWWG/Welcome.html}.

\bibitem{WMAP}
D.~N.~Spergel {\it et al.}  [WMAP Collaboration],
%``First Year Wilkinson Microwave Anisotropy Probe (WMAP) Observations:
%Determination of Cosmological Parameters,'' 
Astrophys.\ J.\ Suppl.\  {\bf 148} (2003) 175
[arXiv:astro-ph/0302209].
%%CITATION = ASTRO-PH 0302209;%%

\bibitem{BBN}
R.~H.~Cyburt, B.~D.~Fields and K.~A.~Olive,
%``Primordial Nucleosynthesis in Light of WMAP,''
Phys.\ Lett.\ B {\bf 567} (2003) 227
[arXiv:astro-ph/0302431].
%%CITATION = ASTRO-PH 0302431;%%

\bibitem{SN}
S.~Perlmutter {\it et al.}  [Supernova Cosmology Project Collaboration],
%``Measurements of Omega and Lambda from 42 High-Redshift Supernovae,''
Astrophys.\ J.\  {\bf 517} (1999) 565
[arXiv:astro-ph/9812133];
%%CITATION = ASTRO-PH 9812133;%%
A.~G.~Riess {\it et al.}  [Supernova Search Team Collaboration],
%``Observational Evidence from Supernovae for an Accelerating Universe and
%a
%Cosmological Constant,''
Astron.\ J.\  {\bf 116} (1998) 1009
[arXiv:astro-ph/9805201].
%%CITATION = ASTRO-PH 9805201;%%

\bibitem{Guth}
A.~H.~Guth,
%``The Inflationary Universe: A Possible Solution To The Horizon And 
%Problems,''
Phys.\ Rev.\ D {\bf 23} (1981) 347.
%%CITATION = PHRVA,D23,347;%%

\bibitem{LSP}
J.~R.~Ellis, J.~S.~Hagelin, D.~V.~Nanopoulos, K.~A.~Olive and
M.~Srednicki,
%``Supersymmetric Relics From The Big Bang,''
Nucl.\ Phys.\ B {\bf 238} (1984) 453.
%%CITATION = NUPHA,B238,453;%%

\bibitem{LHC}
ATLAS Collaboration, {\it Physics Technical Proposal},
{\tt http://atlas.web.cern.ch/Atlas/TP/tp.html};\\
CMS Collaboration, {\tt http://cmsinfo.cern.ch/Welcome.html/}.

\bibitem{newbench}
M.~Battaglia, A.~De Roeck, J.~R.~Ellis, F.~Gianotti, K.~A.~Olive and
L.~Pape,
%``Updated post-WMAP benchmarks for supersymmetry,''
Eur.\ Phys.\ J.\ C {\bf 33} (2004) 273
[arXiv:hep-ph/0306219].
%%CITATION = HEP-PH 0306219;%%

\bibitem{Tibet}
J.~R.~Ellis, K.~A.~Olive, Y.~Santoso and V.~C.~Spanos,
%``Prospects for sparticle discovery in variants of the MSSM,''  
arXiv:hep-ph/0408118.
%%CITATION = HEP-PH 0408118;%%

\bibitem{PS}
Planck Science Team,
{\tt http://www.rssd.esa.int/index.php?project=PLANCK}.

\bibitem{H}
S.~W.~Hawking,
%``The Unpredictability Of Quantum Gravity,''
Commun.\ Math.\ Phys.\  {\bf 87} (1982) 395.
%%CITATION = CMPHA,87,395;%%

\bibitem{EHNS}
J.~R.~Ellis, J.~S.~Hagelin, D.~V.~Nanopoulos and M.~Srednicki,
%``Search For Violations Of Quantum Mechanics,''
Nucl.\ Phys.\ B {\bf 241} (1984) 381.
%%CITATION = NUPHA,B241,381;%%

\bibitem{Witten}
E.~Witten,
%``Quantum gravity in de Sitter space,''
arXiv:hep-th/0106109.
%%CITATION = HEP-TH 0106109;%%

\bibitem{EMNV}
J.~R.~Ellis, N.~E.~Mavromatos and D.~V.~Nanopoulos,
%``Quantum mechanics and black holes in four-dimensional string theory,''
Phys.\ Lett.\ B {\bf 278} (1992) 246
[arXiv:hep-th/9112062];
%%CITATION = HEP-TH 9112062;%%

\bibitem{EMNQMV}
J.~R.~Ellis, N.~E.~Mavromatos and D.~V.~Nanopoulos,
%``String theory modifies quantum mechanics,''
Phys.\ Lett.\ B {\bf 293} (1992) 37
[arXiv:hep-th/9207103].
%%CITATION = HEP-TH 9207103;%%

\bibitem{ABEN}
I.~Antoniadis, C.~Bachas, J.~R.~Ellis and D.~V.~Nanopoulos,
%``An Expanding Universe In String Theory,''
Nucl.\ Phys.\ B {\bf 328} (1989) 117.
%%CITATION = NUPHA,B328,117;%%

\bibitem{EMNI}
J.~R.~Ellis, N.~E.~Mavromatos and D.~V.~Nanopoulos,
%``A String scenario for inflationary cosmology,''
Mod.\ Phys.\ Lett.\ A {\bf 10} (1995) 1685
[arXiv:hep-th/9503162].
%%CITATION = HEP-TH 9503162;%%

\bibitem{EMNS2}
J.~R.~Ellis, N.~E.~Mavromatos, D.~V.~Nanopoulos and A.~Sakharov,
%``Brany Liouville inflation,''
arXiv:gr-qc/0407089.
%%CITATION = GR-QC 0407089;%%

\bibitem{EMNA}
J.~R.~Ellis, N.~E.~Mavromatos and D.~V.~Nanopoulos,
%``String theory and an accelerating universe,''
arXiv:hep-th/0105206.
%%CITATION = HEP-TH 0105206;%%

\bibitem{ELMN}
J.~R.~Ellis, J.~L.~Lopez, N.~E.~Mavromatos and D.~V.~Nanopoulos,
%``Precision tests of CPT symmetry and quantum mechanics in the neutral kaon
%system,''
Phys.\ Rev.\ D {\bf 53} (1996) 3846
[arXiv:hep-ph/9505340];
%%CITATION = HEP-PH 9505340;%%
R.~Adler {\it et al.}  [CPLEAR collaboration],
%``Test of CPT Symmetry and Quantum Mechanics with Experimental data from
%CPLEAR,''
Phys.\ Lett.\ B {\bf 364} (1995) 239
[arXiv:hep-ex/9511001].
%%CITATION = HEP-EX 9511001;%%

\bibitem{Nature}
G.~Amelino-Camelia, J.~R.~Ellis, N.~E.~Mavromatos, D.~V.~Nanopoulos and 
S.~Sarkar,
%``Potential Sensitivity of Gamma-Ray Burster Observations to Wave 
%Dispersion
%in Vacuo,''
Nature {\bf 393} (1998) 763
[arXiv:astro-ph/9712103].
%%CITATION = ASTRO-PH 9712103;%%

\bibitem{EMNS1}
J.~R.~Ellis, N.~E.~Mavromatos, D.~V.~Nanopoulos and A.~S.~Sakharov,
%``Quantum-gravity analysis of gamma-ray bursts using wavelets,''
Astron.\ Astrophys.\  {\bf 402} (2003) 409
[arXiv:astro-ph/0210124].
%%CITATION = ASTRO-PH 0210124;%%

\end{thebibliography}
\end{document}